\documentclass[12pt]{article}
\usepackage{a4wide}
\begin{document}
\thispagestyle{empty}

\begin{flushright}
  THEF-NYM-97.01 \\
  KVI\hspace{2.5mm}1318\\
  nucl-th/9802084
\end{flushright}

\vspace{2\baselineskip}

\begin{center}
  {\large\bf THE STATUS OF THE PION-NUCLEON COUPLING CONSTANT} \\[1.5cm]
  J.J.\ DE SWART$\,^a$, M.C.M.\ RENTMEESTER$\,^b$,
          R.G.E.\ TIMMERMANS$\,^c$ \\ [0.5cm]
  {\it $^a$ University of Nijmegen, Nijmegen, The Netherlands} \\
  {\it $^b$ The Flinders University of South Australia,
  Bedford Park, Australia} \\
  {\it $^c$ KVI, University of Groningen, Groningen, The Netherlands}
\end{center}

\vspace{1cm}

\begin{center}
  ABSTRACT
\end{center}

A review is given of the various determinations of the different
$\pi N\!N $ coupling constants in analyses of the low-energy
$pp$, $np$, $\overline{p}p$, and $\pi p$ scattering data.
The most accurate determinations are in the energy-dependent
partial-wave analyses of the $N\!N$ data.
The recommended value is $f^2=0.075$.
A recent determination of $f^2$ by the Uppsala group from
backward $np$ cross sections is shown to be model dependent
and inaccurate, and therefore completely uninteresting.
We also argue that an accurate determination of $f^2$ using
$pp$ forward dispersion relations is not a realistic option.

\vfill

\begin{center}
  Invited talks given by J.J.\ de Swart at \\
  FBXV, the 15th Int. Conf. on Few Body Problems in Physics,\\
  held from July 22 to 26, 1997 in Groningen, The Netherlands, and at \\
  MENU97, the 7th Int. Symp. on Meson-Nucleon Physics,\\
  held from July 28 to August 1, 1997 in Vancouver, B.C., Canada. \\
  Published in  TRIUMF Report TRI-97-1, 96 (1997).
\end{center}

\newpage
\pagenumbering{arabic}

\section{Introduction}

It is exactly 50 years ago that the pion was discovered in cosmic
rays by Lattes, Occhialini, and Powell~\cite{La47} in Bristol.
Since then we have come to know quite a lot about the pion and
its interactions. We will concentrate here on the
pion-nucleon-nucleon $(\pi N\!N)$ coupling constants,
and review the various determinations of the
$\pi N\!N$ coupling constants in analyses of the low-energy
$pp$, $np$, $\overline{p}p$, and $\pi p$ scattering data.

One can distinguish in principle 4 different $\pi N\!N$
coupling constants. The coupling constant
$ f(p\pi^- \rightarrow n) = \sqrt{2} \, f_{-} $
denotes the strength at the vertex at which
a proton is annihilated and a neutron created and where either
a $\pi^{-}$ is annihilated or a $\pi^{+}$ created.
The coupling constant
$ f(n\pi^+ \rightarrow p) = \sqrt{2} \, f_{+} $
denotes the strength at the vertex at which
a neutron is annihilated and a proton created and where either
a $\pi^{+}$ is annihilated or a $\pi^{-}$ created.
The coupling constant
$ f(p\pi^0 \rightarrow p) = \: f_{p} $
denotes the strength at the vertex at which
a proton is annihilated and another proton created and where
a $\pi^{0}$ is either annihilated or created.
The coupling constant
$ f(n\pi^0 \rightarrow n) = \: - \: f_{n} $
denotes the strength at the vertex at which
a neutron is annihilated and another neutron created and where
a $\pi^{0}$ is either annihilated or created.
Important combinations are $f_{c}^{2} = f_{+} \; f_{-} $ and
$ f_{0}^{2} = f_{p} \; f_{n} $.
The factors in front of $ f_{+} \: , f_{-} \: , $ and $ \: f_{n} $
are chosen in such a way that we get in the case of
{\it charge independence}:
\begin{equation}
   f_{p} = f_{n} = f_{+} = f_{-} = f \ .
\end{equation}

Because the strong interactions are invariant under charge
conjugation there is a relation between the pion-nucleon-nucleon
coupling constants and the pion-antinucleon-antinucleon coupling
constants. By using the charge-conjugation operator $C$ we can
define the antiproton $\overline{p}=Cp$ and the antineutron
$\overline{n}=Cn$. The neutral pion is its own antiparticle
and is an eigenstate of the charge-conjugation operator
$C\pi^{0} = \eta_{c}\pi^{0}$
with as eigenvalue the charge parity $\eta_{c}=1$ .
The coupling constants of the neutral pion to the antinucleons
are related to the coupling constants of the neutral pion
to the nucleons by
\begin{equation}
  f(\overline{p}\pi^{0}\rightarrow \overline{p}) =
  \eta_{c} f(p\pi^{0}\rightarrow p) \quad {\rm and} \quad
  f(\overline{n}\pi^{0}\rightarrow \overline{n}) =
  \eta_{c} f(n\pi^{0}\rightarrow n)  \ .
\end{equation}
This leads to the following relations
\begin{eqnarray}
\begin{array}{lcl}
 f(\overline{p}\pi^{0}\rightarrow\overline{p}) = \hspace{4mm} f_{p} & , &
 f(\overline{p}\pi^{+}\rightarrow\overline{n}) = \sqrt{2}f_{-} \\
 f(\overline{n}\pi^{0}\rightarrow\overline{n}) = - f_{n} & , &
 f(\overline{n}\pi^{-}\rightarrow\overline{p}) =  \sqrt{2}f_{+} \ .
\end{array}
\end{eqnarray}

The pion is a pseudoscalar $(J^{PC} = 0^{-+})$ particle. The
coupling of the pion to the nucleons can be described by either the
PS (pseudoscalar) interaction lagrangian ${\cal L}_{PS}$ or the
PV (pseudovector) interaction lagrangian ${\cal L}_{PV}$, where
\begin{equation}
  {\cal L}_{PS} = g\sqrt{4\pi}(i\overline{\psi}\gamma_{5}\psi)\phi
  \quad {\rm and} \quad {\cal L}_{PV} = \frac{f}{m_{s}}\sqrt{4\pi}
  (i\overline{\psi}\gamma_{\mu}\gamma_{5}\psi)\partial^{\mu}\phi \ .
\end{equation}
We prefer to use the PV coupling, because of chiral symmetry and
because of SU(3,$F$) flavor symmetry of the pseudoscalar-meson
couplings to the baryon octet~\cite{Ti90}.
The scaling mass $m_{s}$ was introduced to make the PV
coupling constant dimensionless. It is convention~\cite{Eb71} to
choose $m_{s}=m_{+}$, where $m_{+}$ is the mass of the $\pi^{+}$.
Unfortunately, there are physicists~\cite{Er83}
who take $m_{s}$ equal to the mass of the quanta of the field
$\phi$. This leads to an unnecessary large charge-independence
breaking in the coupling constants. We strongly advice to use
the convention. The PS coupling constant $g$ and the PV coupling
constant $f$ are related by~\cite{Dy48} the {\it equivalence relation}
\begin{equation}
    \frac{g}{M_1 + M_2} = \frac{f}{m_s}  \ .
\end{equation}
This means $g^2_p = 180.773 f^2_p$, $g^2_c = 181.022 f^2_c$,
and $g^2_n = 181.271 f^2_n$.

One must realize that charge independence or SU(2,$I$) isospin
symmetry for the PV coupling constants leads to broken SU(2,$I$)
symmetry for the PS coupling constants. For example, when we take
$f^{2} = 0.0747$, then we get
$g_{p}^{2} = 13.50$, $g_{+}^{2} = g_{-}^{2} = 13.52$,
and $g_{n}^{2} = 13.54$.
For most purposes charge independence for the values of
the coupling constants can be assumed. We recommend to use
\begin{equation}
     {\rm either} \quad f^{2} = 0.075 \ , \quad
     {\rm or}     \quad g^{2} = 13.5  \ .
\end{equation}
This implies that the Goldberger-Treiman discrepancy
is only $\Delta_{\pi N}\simeq 2\%$.
\vspace{\baselineskip}

\section{Early determinations}

It is impossible for us to give a proper and historically correct
review of the various determinations of $f^{2}$.
To get an impression about what was going on we will give a short,
incomplete, and at certain places possibly incorrect, description.

In the early fifties pion-photoproduction experiments were much
more accurate than the $\pi N$ scattering experiments.
Also the $N\!N$ scattering experiments were still in their infancy.
To our knowledge, the first estimate $f_c^2\simeq 0.3$
of the $\pi N\!N$ coupling constant
came in 1950 from the photoproduction of charged pions~\cite{As50}.
The Kroll-Ruderman Theorem~\cite{Kr54} shows that $f^2_c$ can be
determined directly from the photoproduction reactions near threshold.
Using this method Bernardini and Goldwasser~\cite{Be54} found in 1954
the already much better value $f_c^2=0.065$.

Perhaps the best place to determine the $\pi N\!N $ coupling
constants is from the $N\!N$ scattering data.
Probably the first determination in $N\!N$ scattering
was made in 1952 by M.M. L\'{e}vy~\cite{Le52}. In order to
reproduce the low-energy parameters (binding energy of the
deuteron and the $s$-wave scattering lengths) he needed $f^{2} = 0.054$.
In 1955 Chew and Low's static nucleon extended-source interaction
Hamiltonian (PV) was applied by Gartenhaus~\cite{Ga55} in his
calculation of the one- and two-pion-exchange $N\!N$ potential.
This potential with $f^2$ = 0.089 described the low-energy
$N\!N$ data well.

Encouraged by L\'{e}vy's result
Chew~\cite{Ch54} studied around 1953 in a series of papers the
low-energy $\pi N$ scattering data using the nonrelativistic
Tamm-Dancoff method. He found $ f_{c}^{2} = 0.058 $.
At about the same time Sartori and Watighin~\cite{Sa54}
used the Cini-Fubini method to study the same data.
They came up with the value $ f_{c}^{2} = 0.065 $.

The very first applications of dispersion relations to the $\pi N$
data to determine the pion-nucleon coupling were by
Haber-Schaim~\cite{Ha56}, Davidon and Goldberger~\cite{Da56},
and Gilbert~\cite{Gi57}.
The value obtained by Haber-Schaim was $f_{c}^{2} = 0.08(1)$.
Using forward dispersion relations Schnitzer and Salzman~\cite{Sc59}
obtained in a careful analysis of the $\pi N$ data the value
$f^{2} = 0.08(1)$.

Then there is
the determination of $f_{c}^{2}$ by Jackson~\cite{Ja58}.
Using the Chew-Low effective-range plot for the
$P_{33}$ phase shift he found $f_{c}^{2} = 0.08(2)$.

In 1958 Chew~\cite{Ch58} pointed out that the backward $np$
scattering was possibly a good place to determine $f_{c}^{2}$.
As will be pointed out later, that ``this is a good place'' has
{\em incorrectly} become part of our folklore. This method was
for the first time applied to the $np$ data by Cziffra and
Moravcsik~\cite{Cz59}
with the result that $f_{c}^{2}$ is between 0.06 and 0.07
with a very large error.
They already pointed out that a partial-wave analysis of $N\!N$
scattering data would not suffer from the disadvantages of an
extrapolation method and give more reliable and realistic results.

Around this time the outstanding series of $N\!N$ partial-wave
analyses (PWA) by the Livermore-Berkeley group was started. They
regularly produced a value for the $\pi N\!N$ coupling constant.
In a single-energy phase-shift analysis of the $pp$ data
at 310 MeV MacGregor {\it et al.}~\cite{Ma59} found in 1959
either $f_{p}^{2} = 0.062(11)$ or $f_{p}^{2} = 0.069(17)$.
In 1968 they~\cite{Ma68} found for this coupling constant
$f_{p}^{2} = 0.081(5)$.
Applying forward dispersion relations to the $pp$ data
Bugg~\cite{Bu68}
determined in 1968 the value $f_{p}^{2} = 0.075(4)$.

In a study of the $\pi N$ data in 1973 Bugg {\it et al.}~\cite{Bu73}
used fixed-$t$ dispersion relations to
determine the coupling constant $f_{c}^{2} = 0.079(1)$.
They stated over-optimistically that their small error is
``believed to cover both statistical and systematic uncertainties.''
This determination has been the standard for almost two decades.

The outstanding Karlsruhe-Helsinki partial-wave analyses~\cite{Ho83} of
the $\pi N$ data used the value of Bugg {\it et al.} as input. In 1980
Koch and Pietarinen~\cite{Ko80} used fixed-$t$ dispersion relations
and found again that $f_{c}^{2} = 0.079(1)$.
However, this is more a consistency check than a real determination,
because the value of the coupling constant was used as
{\it input} in the analyses. Other values of $f^2_c$ were
not tried as input. It is interesting to read the claim
made by Koch and Pietarinen that ``the error is our estimate due to
systematic effects. The statistical error is completely negligible''!
It is incomprehensible to us how theory can remove or reduce
the statistical errors. It should also be remarked that the
Karlsruhe-Helsinki analyses of the $\pi N$ data did {\em not}
include a treatment of the normalization errors of experimental data.

In 1981 Kroll~\cite{Kr81} applied $pp$ forward dispersion relations
to determine $f_{p}^{2}$. He came to the conclusion that
$f_{p}^{2} = 0.080(2)$. Later on, more attention will be paid to
this special method.

We see that around 1980 all dispersion-relation determinations
from $\pi N$ data as well as from $N\!N$ data seem to agree on
the charge-independent value $f^{2} = 0.079(1)$.
Dispersion relations were then considered by many people {\em the}
method to be used in determining coupling constants.

For references to other work on the determination of
coupling constants the reader is referred to 
the classic book~\cite{Be55} by H.A.\ Bethe and F.\ de Hoffmann and to
the various editions~\cite{Eb71,Eb70,Pi73,Na76,Na79,Du83} of the
``Compilation of Coupling Constants and Low-Energy Parameters.''
\vspace{\baselineskip}

\section{Newer determinations}

In 1975 the Nijmegen group~\cite{Na75} published the
Nijmegen D $N\!N$ potential.
This is a hard-core potential, which reproduced the then available
$N\!N$ data very well. It required $f^{2} = 0.0741$.
It is interesting to see that already then the $N\!N$ data favored
a small value for the coupling constant and that the deuteron
properties of that potential were excellent. The soft-core
Nijmegen $N\!N$ potential Nijm78 was published~\cite{Na78} in 1978.
Also in this potential the coupling constant preferred
a low value. Because we were at that time still indoctrinated by
the $ \pi N $ people that $f^{2} = 0.079(1)$, we prevented
our coupling constant from going ``too low,'' and ended up
with the value $f^{2} = 0.0772$.

It was around 1983 that we became convinced that the $\pi^{0}pp$
coupling constant was much lower than 0.079. At the 1983 Few-Body
Conference in Karlsruhe we~\cite{Sw84} stated that ``we believe
that $f_{p}^{2}$ is probably more in the neighborhood of 0.075
than of 0.080.''
This belief was based on our PWA of the low-energy $pp$ data.
It lasted until 1987 before we could make our suspicions hard,
that is, before we could give solid evidence based on statistics.
In our energy-dependent PWA~\cite{Be87} of the $pp$ scattering data
below $T_{lab} = 350$ MeV  we found $ f_{p}^{2} = 0.0725(6)$.
Unfortunately, the magnetic-moment interactions were not yet included
in these analyses. When that was finally done~\cite{St90},
we~\cite{Be90,Ri90} obtained $f_{p}^{2} = 0.0749(6)$. At that
time we had not made a determination of $f_c^2$ by ourselves.
Therefore we had to assume that the value for $f^2_c$ determined
in $\pi N$ scattering was correct and that there was thus evidence
for a large breaking of charge independence. That raised
hell. Many people~\cite{Th89,Ho90,Co90,Ma91,Er92,Bu92,Ha92}
had their own explanation about what was wrong with {\it our}
analyses. Those ``explanations'' were almost all based on wrong
assumptions and ignorance about energy-dependent PWA's.
Nobody, on the other hand, questioned seriously
the value for $f_c^2$ from $\pi N$ analyses.

Of course there are people who are familiar with PWA's.
Arndt {\it et al.} noticed that in the energy-dependent VPI\&SU
analyses of the $np$ data a lower coupling constant was favored.
This encouraged them to have a fresh look at the $\pi N$ data.
In 1990 the VPI\&SU group~\cite{Ar90} came up with the new value
$f_c^2 = 0.0735(15)$. Again, many
people~\cite{Er92,Ho91,Lo92,St92,Bu93}
had their own ``explanation'' about what was wrong with
{\em this} analysis. Of course, nothing was really wrong!
In order to try to appease the opposition the VPI\&SU
group~\cite{Ar94} included
dispersion-relation constraints in their analyses of the $\pi N$
data. This lead then to the value $f_c^2 = 0.0761(8)$.
In our opinion, it is better not to impose such constraints,
because it is almost impossible to make the proper corrections
for the electromagnetic interactions.

In Nijmegen we were at that time also very busy to get determinations
of $f_{c}^{2}$. A study of the backward $np$ data was very
disappointing. Despite the folklore~\cite{Ch58} that this
charge-exchange reaction was the best place to
determine $f_{c}^{2}$, we were unable to get a sufficiently accurate
determination~\cite{Re91} such that we could distinguish between
the values 0.075 and 0.079 for $f_{c}^{2}$.

An energy-dependent analysis~\cite{Ti91} in 1991 of the charge-exchange
reaction $\overline{p}+p \rightarrow \overline{n}+n$ gave us the
value $f_{c}^{2} = 0.0751(17)$. To convince ourselves that we were
really looking at the OPE potential, we determined the mass of the
exchanged particle. This turned out to be $m_{\pi^+} = 145(5)$ MeV,
which must be compared with the experimental value
$m_{\pi^+} = 139.57$ MeV. This 1991 result was the main motivation
for a new accurate measurement of the charge-exchange cross section
by the PS206 group at LEAR. The completed $\overline{N}\!N$ PWA of
1993 gave~\cite{Ti94} $f_c^2=0.0732(11)$.

Our energy-dependent PWA's of the $np$ data
required an enormous effort.
Finally, in 1991 our first analysis~\cite{Kl91} of the combined
$pp$ and $np$ data was completed. We then were able to make independent
determinations of various $\pi N\!N$ coupling constants. We found
$f_{p}^{2} = 0.0751(6)$, $f_{0}^{2} = 0.0752(8)$, and
$f_{c}^{2} = 0.0741(5)$. For the charge-independent value
we found $f^{2} = 0.0749(4)$.

It was now evident that the energy-dependent PWA's of the
$pp$, $np$, $\overline{p}p$, and $\pi p$ data all lead to
about the same value $f^2 = 0.075$ for the $\pi N\!N$ coupling
constant. We were surprised to note the many physicists trying
to hold on to the old values.
To counter their arguments we published in
1993 a paper~\cite{St93} in which we studied in detail
various aspects of our determinations of coupling constants.
In that paper~\cite{St93} we paid special attention
to Chew's proposal~\cite{Ch58} to determine $f^2_c$ from
single-energy studies of the backward $np$ data.
We showed explicitly that extrapolating to the pion pole is
{\em very inaccurate}, and clearly not a good method to make
a reliable determination of $f_{c}^{2}$ with a small error.
\vspace{\baselineskip}

\section{Present situation}

The coupling constant $f^2_p$ can best be determined from the
$pp$ scattering data. Our latest published value stems from 1993.
In an analysis~\cite{St93a} of 1787 $pp$ data below 350 MeV with
21 model parameters, we obtained $\chi ^{2}_{min} = 1787$ and
$f_{p}^{2} = 0.0745(6)$.

This analysis has been updated~\cite{Re97}. The 1997 Nijmegen $pp$
database contains 1955 data. These can be described very well
with 20 model parameters.
We find $\chi _{min}^{2} = 1962$ and $f_{p}^{2}$ = 0.0753(5).
When we leave out the recent
data from Haeberli {\it et al.}~\cite{Ha97},
then we get $f_{p}^{2}$ = 0.0747(5). We have here a nice example
of a {\em non}-dedicated experiment, which {\em does} influence the
value of the coupling constant.

The $np$ scattering data give the combinations
$f_{c}^{2}$ and $f_{0}^{2}$. In the 1993 Nijmegen PWA~\cite{St93a}
of the $pp$ and $np$ data below 350 MeV we analyzed
2512 $np$ data. In this analysis we found
$\chi _{min}^{2} = 2480$ and $f_{c}^{2} = 0.0748(3)$.
We have also been able to do PWA's of the $np$ data alone~\cite{Kl94}.
This analysis uses all $np$ data, $N_{data}$ = 3964,
below 500 MeV. We get $\chi _{min}^{2} = 4005$ with
$f_{c}^{2} = 0.0748(3)$ and $f_{0}^{2} = 0.0745(9)$.

Another useful place to determine these coupling constants is in
$\overline{N}\!N$ scattering.
The elastic $\overline{p}p$ data give $f_{p}^{2}$,
and the charge-exchange data give $f_{c}^{2}$.
In these analyses~\cite{Ti91,Ti94,Ti97} all data
below 925 MeV/c were analyzed.
In 1991 884 observables were analyzed~\cite{Ti91}. Using
23 model parameters we reached $\chi^{2}_{min}/N_{obs} = 1.15$
and we determined $f_{c}^{2} = 0.0751(17)$.
The 1993 analysis~\cite{Ti94} contained 3646 data (elastic as
well as charge exchange) and 30 model parameters were used;
$\chi^{2}_{min}/N_{data} = 1.04$ was reached with
$f^{2}_{c} = 0.0732(11)$.
This analysis has been updated, where the latest charge-exchange
data from the LEAR experiments PS199~\cite{La95,Ah95} and
PS206~\cite{Bi94} were incorporated. Analyzing now 3847 data with
36 model parameters~\cite{Ti97} $\chi^{2}_{min}/N_{data} = 1.05$
was obtained with $f_{c}^{2} = 0.0736(10)$.

Traditionally, the most important source for the $\pi N\!N$ coupling
constant has been $\pi N$ scattering. Neutron exchange in elastic
$\pi^{+}p$ scattering depends on the combination $f_{c}^{2}$.
The analysis of the elastic and charge-exchange $\pi^{-}p$
scattering is more complicated, and various combinations of the
coupling constants come into play. We have already seen that
Arndt {\it et al.} determined in 1990~\cite{Ar90} the value
$f^{2}_{c}$ = 0.0735(15). In 1994 they~\cite{Ar94} included
dispersion-relation constraints. This raised their value for
$f^2_c$ to 0.0761(8). Bugg has redone his 1973 dispersion
analysis~\cite{Ma93} and his updated value is
$f^{2}_{c} = 0.0771(14)$. This determination is {\it not}
based on an energy-dependent PWA.

Another analysis is the Los Alamos-Groningen energy-dependent
PWA of the $\pi N$ data by Timmermans~\cite{Ti98}.
After a careful selection the
$\pi^{+}p$ database below 410 MeV contains
1092 data. The PWA uses 14 parameters and reaches
$\chi^{2}_{min}/N_{data} = 0.98$ with $f^{2}_{c} = 0.0743(8)$.
Extending this analysis to the coupled $\pi ^{-}p$ and $\pi^{0}n$
channels results in a database containing 898 data.
In this analysis 17 parameters were used and
$\chi^{2}_{min}/N_{data} = 1.11$ was reached with the
charge-independent coupling $f^2 = 0.0756(9)$.

It takes too much space and time for a proper discussion of all
determinations
of the coupling constants in single-energy PWA's. Such determinations
are inherently much more inaccurate than the energy-dependent
analyses. One of the reasons is just the number of data that
is involved. In single-energy PWA's this is about a few tens,
but in energy-dependent PWA's about a few thousands.
Roughly a factor of 100 difference. This implies about a factor of 10
difference in the statistical accuracy.
Because we feel very strongly that single-energy PWA's
are vastly inferior to energy-dependent PWA's, we will not
discuss determinations based on such single-energy studies
(with the exception of the recent Uppsala work, see below).

It must be clear from the above, that ($i$) the energy-dependent
PWA's of the about 12.000 $N\!N$, $\overline{N}\!N,$ and $\pi N$
data are all in agreement, that ($ii$) there is at present no
evidence for a breaking of
charge independence in the coupling constants, and that ($iii$)
the charge-independent value is $f^2 \simeq 0.0750$.

\section{Breaking of charge independence}

In which way the breaking of charge independence influences
the values of the coupling constants is
a widely studied topic~\cite{Mo68,Th81,He87,Fr96,Me96,Hw96,Ma97}.
An earlier paper discussing the radiative corrections was by
Morrison~\cite{Mo68}. Later,
estimates were made using various quark models~\cite{Th81,He87}
and more recently QCD sum rules~\cite{Me96} were applied.
A difficulty with these various methods and models is that the
predictions differ (even in sign) from model to model.
In some of the simpler models one has the relation
$ f_{c} = (f_{p} + f_{n})/2 $.
Friar, Goldman, and van Kolck studied this issue in chiral
perturbation theory~\cite{Fr96}.

Using the QCD sum rules for the appropriate pion-nucleon
three-point functions Meissner and Henley~\cite{Me96} calculated
the splitting between $f_{p}$, $f_{n}$, and $f_{c}$
due to isospin breaking in the strong interaction. They found
a large breaking of isospin in these coupling constants,
the lower limit of which will result approximately in
$f_{n}^{2}$ = 0.074, $f_{c}^{2}$ = 0.075,
and $f_{p}^{2}$ = 0.076.

The present accuracies in the determination of the various coupling
constants are such, that with a little improvement in the data and
in the analyses these charge-independence breaking effects could
be checked.
\vspace{\baselineskip}

\section{Energy-dependent PWA versus \protect\newline single-energy PWA}

In the energy region below T$_{lab}=350$ MeV {\em one}
energy-dependent PWA of all $N\!N$ data could be performed.
Let us take as example the Nijmegen 1993 PWA~\cite{St93a}.
We simultaneously analyzed 1787 $pp$ data and 2514 $np$ data for
a grand total of 4301 $N\!N$ data.
Or one could divide all these data in about 10 energy intervals,
clustered around the energies: 0, 1, 5, 10, 25, 50, 100, 150, 210,
and 320 MeV, and perform 10 single-energy PWA's.
Both these total analyses will furnish phase shifts at these
above energies.
The big advantage of the energy-dependent PWA is that the
multienergy (m.e.) phases are much smoother as a function of energy.
The statistical fluctuations in the single-energy (s.e.) phases are
averaged out in the m.e. phases.
As a result, the errors in the m.e. phases are much smaller than
the errors in the s.e. phases.
When new data are added to the database these m.e. phases
are much more stable than the s.e. phases.
What is needed for a successful energy-dependent PWA is a good
description of the energy dependence of the phases.

The difficulty with the single-energy PWA's is that they are
{\em overparametrized}. In the above mentioned m.e. PWA we use 39
parameters. To describe exactly the same data in the 10 s.e. PWA's
we need 116 parameters. In both analyses we use 4301 $N\!N$ data.
In the m.e. analysis we reach $\chi^2_{min} = 4276$.
In the s.e. analyses we describe all data with $\chi^2_{min} = 4096$.
We need 77 more parameters to obtain a drop in $\chi^2$ of
{\em only} 180. This overparametrization of the s.e. phases results
in a large noise content in these phases. This makes such s.e. phases
rather useless. This is the reason that the Nijmegen group will
{\em not} present any more s.e. phases. We strongly feel that such
phases should not be used anymore.

Important in energy-dependent PWA's is a good description of the
fast energy dependence of the phases. The slow energy dependence
can easily be parametrized.
The fast energy dependence is determined by the longest-range forces.
It is therefore important to have a good description of the
long-range electromagnetic forces. In the Nijmegen analyses we have
paid very special attention~\cite{Au83,St90}
to these longest-range forces.
The longest-range nuclear forces are due to one-pion exchange (OPE).
An important part of the energy dependence of the phases comes
from OPE. This is the main reason that we can determine the
coupling constants so accurately.
The most accurate data are the $pp$ data. The $np$ data are definitely
less accurate. Still, we can determine the charged-pion
coupling constant
$f^2_c$ in the $np$ PWA with the very small error $3\times 10^{-4}$.
The reason is that in $np$ scattering the electromagnetic force is
essentially absent (only magnetic-moment interactions).
The long-range OPE force in $pp$ scattering is hidden under the
longer-range electromagnetic interaction. This ``explains'' why
we can determine $f^2_p$ only with the somewhat larger error
$6\times 10^{-4}$.
\vspace{\baselineskip}

\section{Backward np scattering}

Recently, backward $np$ scattering experiments at $T_{lab} = 162$
MeV were performed in Uppsala~\cite{Er95}. These data were then used
in a modified Chew extrapolation method to determine $f^2_c$.
The authors claim a ``high'' value for $f^2_c$ with an incredible small
error. This value is obtained from only 31 data and is in flagrant
disagreement with the values deduced in the energy-dependent PWA's of
the 12.000 $pp$, $np$, $\overline{N}\!N$, and $\pi N$ data.
Let us look at what is going on.

\subsection{OPE amplitude}
Forgetting about the $np$ mass difference the one-pion-exchange
amplitude evaluated in Born approximation reads
\begin{equation}
   M_{OPE} = - \frac{4\pi f^2}{m^2_s} \
    \frac{ (\mbox{\boldmath $\sigma$}_1\cdot\mbox{\boldmath $k$})
           (\mbox{\boldmath $\sigma$}_2\cdot\mbox{\boldmath $k$}) }
     {k^2 + m_\pi^2} \ ,
\end{equation}
with $ k^2 = 2p^2 (1\mp\cos\theta)$. Here $p$ is the
center-of-mass momentum. We see that $M_{OPE}$ is
energy ($E$ or $p^{2}$) dependent, $\theta$ dependent, and
mass dependent.
The Nijmegen energy-dependent PWA's take {\em both} the energy
dependence and the $\theta$ dependence into account. Also the
mass dependence has been studied and the masses of the exchanged
particles were determined with excellent results~\cite{Kl91}.

Recently, there has appeared a new gimmick: {\em dedicated}
experiments. This seems to mean: \\
   ``{\it Dedicated to the proposition that all PAC's
          are equally easily led astray.\/}'' \\
In the analyses of the dedicated experiment~\cite{Er95}
{\it only} the $\theta$ dependence of the one-pion-exchange
amplitude is taken into account, and that only approximately.

We would like to point out that the $\pi^{0}$-exchange amplitude
vanishes in the forward direction ($\theta$ = 0), and that the
$\pi^{\pm}$-exchange amplitude (forgetting the $np$ mass difference)
vanishes at $\theta = 180^\circ$. This implies that the backward
peak ($\theta = 180^\circ$) in $np$ scattering is {\it not}
due to $\pi^{\pm}$ exchange. The fall-off of the differential cross
section for $\theta < 180^\circ$ is due to destructive interference
between $\pi^{\pm}$ exchange and the background.

\subsection{The data}
The data, as published~\cite{Er95}, have an incredibly large $\chi^2$
with respect to the standard PWA's of the VPI\&SU~\cite{SAID}
and Nijmegen~\cite{NOL} groups. We find $\chi^2=291.6$ for 31 data.
We have tried to understand what causes this.
The data consist of several independent and unnormalized data sets.
When one compares these sets in the regions where they overlap
one notices inconsistencies.
Next, these sets are {\em incorrectly} normalized.
In this way large systematic errors are introduced.
Our conclusion about these data is that,
first, one must clean up the systematic deviations between the
individual data sets and that,
second, the data badly need reanalysis, where better attention
should be paid to the normalization.

We want to point out that the problem is not just between
the Uppsala data and other backward $np$ cross sections.
Even if {\it all} backward $np$ cross sections are removed
from the database, the Uppsala data are still in flagrant
disagreement with the remainder of the data. This means that
the Uppsala data are also in disagreement with all the spin
observables, such as polarization and spin-correlation data.
Moreover, as mentioned, the Uppsala data are internally inconsistent.

The Uppsala group has claimed that there are two ``families''
of $np$ cross sections, and that in the Nijmegen PWA only one
``family'' is accepted. This unfounded claim we find ludicrous.
According to our findings, the second ``family'' consists {\it only}
of the incorrectly normalized Uppsala data.

\subsection{Extrapolation Method}
The authors try to extrapolate the backward differential cross section
to the charged-pion pole. They consider several extrapolation
procedures and claim that their modified Chew extrapolation method
is model independent. Moreover, they think that this method can reduce
the statistical errors. We would like to stress that {\it no method
of analysis can reduce statistical errors\/}: These are inherent to
the experimental data.

Arndt {\it et al.}~\cite{Ar95} used exactly the same method as
used by the Uppsala group to {\em all} available backward $np$
data sets. They showed
that the {\em average} of all extrapolations is consistent with
$f^2_c = 0.075$ and that the error in each individual determination
is much larger (about a factor of 10) than
claimed by the Uppsala group.

The model dependence of the Uppsala method is easily demonstrated.
We applied exactly the same method as used by the Uppsala group. For
the comparison potential we use the Nijmegen $np$ potential Nijm93,
as was done by the Uppsala group. However, we constructed different
versions by only changing the $\pi N\!N$ coupling constant $f^2_c$ in
this comparison potential.

In Model I we use $f^2_c$ = 0.075. In the fit to the data 2 parameters
are needed and we find $f^2_c$ = 0.0809(3). The error is here the
estimate of the statistical extrapolation error only.
This is exactly the same model as used by the Uppsala group and our
results are therefore exactly the same as the results of the Uppsala
group.

In Model II we use $f^2_c$ = 0.079. In the fit to the data we now need
4 parameters and we find $f^2_c$ = 0.0808(30). This model
gives the same value for $f^2_c$ as Model I, but the error estimate is
now a factor 10 larger (and much more realistic).

In Model III we use $f^2_c$ = 0.071. To produce a good fit only
1 parameter is needed. This time we find a much lower value for
the coupling constant, $f^2_c$ = 0.0747(1), but
a very unrealistic small error (because in this model we need only 1
parameter: the more parameters, the larger the error).

This simple exercise clearly demonstrates the model dependence
introduced by the comparison potential. We have shown here that both
the value of the coupling constant and the value of the error are
model dependent. We claim that the real error is 0.003 and {\it not}
0.0003 as given by the Uppsala group. The result then is in
total agreement with our previous conclusions~\cite{St93} and
also in agreement with Arndt's results~\cite{Ar95}.

This large error makes the determination of the coupling constant
by the Uppsala group {\em totally uninteresting} and shows that
the label ``dedicated'' for such experiments is presumptuous and
completely unwarranted.

Similar conclusions hold for Chew-type extrapolations of the
forward \\ $\overline{p}+p\rightarrow\overline{n}+n$ cross sections
to determine $f_c^2$~\cite{Br95a,Br95b,Er97}. Such extrapolations
cannot compete with determinations in the energy-dependent
PWA of all $\overline{p}p$ data~\cite{Ti91,Ti94,Ti97}.

\section{Forward dispersion relations}

Forward dispersion relations are often seen as the ultimate
tool in determining coupling constants.
This is definitely not correct for the $pp$ forward dispersion
relations. One reason for this is the problematic treatment
of the electromagnetic interactions~\cite{Gr77,Gr78,Gr80,Kr81}.
The scattering amplitude $M$ between charged particles can be
written as
\begin{equation}
  M = M_{em} + M_{N} \ , \label{eq:EM}
\end{equation}
where $M_{em}$ is the electromagnetic amplitude~\cite{Be90}
and $M_N$ the nuclear amplitude.
Forward dispersion relations cannot be written down for this
nuclear amplitude, because it contains too many remnants of
the electromagnetic interaction. For example,
this amplitude $M_N$ does not have a pion pole, but it has
a branchpoint singularity like $1/(k^2+m_{\pi}^2)^{1+i\eta}$,
where $\eta = \alpha / v_{lab}$ is the
relativistic Rutherford parameter~\cite{Br55}.
By removing all Coulomb effects from $M_N$ one can
construct a new amplitude $M^{\,\prime}_N$, for which one
could possibly write down forward dispersion relations.

A somewhat similar situation exists for the $^1S_0$ $pp$ scattering
length. Here one also likes to remove all Coulomb effects.
This time from the $pp$ scattering length~\cite{Be88}
$a_c(pp) = -7.806(3)$ fm,
such that the resulting nuclear $pp$ scattering length
$a_N(pp)\simeq - 17.3$ fm can be compared to the $np$
scattering length~\cite{St93a} $a(np) = - 23.738$ fm.
The latter is obtained at very low energies
from the effective-range function $F(p^2) = p\cot\delta$
by the zero-range expansion $F(p^2) = - 1/a$.
We will compare $a_N(pp)$ with $M^{\,\prime}_N$.
The $pp$ scattering length $a_{em}(pp)$ which is obtained
from a much more complicated effective-range
function~\cite{Br36,La44,He60,Au82} (including effects
of the full electromagnetic interaction) must then be
compared with $M_N$.
Various $pp$ scattering lengths, $a_c$, $a_E$, and $a_{em}$, exist.
They all arise from different treatments of the
electromagnetic interaction (including vacuum polarization,
relativistic effects, etc.) in the construction of effective-range
functions. We will explain below that there are
also different definitions of $M_N$.

Very important in the treatments of forward dispersion relations
has been the optical theorem.
However, big question marks must be placed by the application of
the optical theorem in reactions between charged particles.
Because of the presence of the long-range Coulomb interaction (with
screening effects neglected) the total cross section $\sigma_T$
is infinite.
This is a well-known result, but the consequences are sometimes not
properly understood~\cite{Bu97,Ho97}. To get finite results one
has to remove in one way or another the Coulomb interaction.
That this is a far from trivial and model-dependent procedure is
often not appreciated. One then defines such theoretical concepts
as the ``total hadronic cross section'' $\sigma^h_{T}$.
This is {\it not} a measurable
experimental quantity, but a theorist's invention. Moreover,
one has to realize that the hadronic cross section is related
to $M_N$ and {\it not} to $M^{\,\prime}_N$. Therefore the hadronic
cross section is {\it not} related to ${\rm Im}\,M^{\,\prime}_N$
and cannot be used in the dispersion integrals. Hence, the
optical theorem cannot be used in connection with $M^{\,\prime}_N$.

The scattering amplitude $M$ between
charged particles was written down in Eq.~(\ref{eq:EM}), where
the electromagnetic amplitude $M_{em}$ is given by~\cite{Be90}
\begin{equation}
   M_{em} = M_{c_1} + M_{c_2} +
                      M_{mm} + M_{vp} \ .
\end{equation}
Here $M_{c_1}$ is the standard Coulomb amplitude, $M_{c_2}$ a
two-photon exchange contribution, $M_{mm}$ the magnetic-moment
contribution, and $M_{vp}$ the vacuum-polarization contribution.
The standard Coulomb amplitude $M_{c_1}$ can be written either as
the divergent partial-wave series~\cite{Ta74,Se75,Ge76}
\begin{equation}
   M_{c_1} =  \sum_{\ell=0}^\infty
                       (2\ell+1)\ \frac{e^{2i\xi_\ell}-1}
                       {2ip}\, P_\ell(x) \ ,
  \label{eq:coul}
\end{equation}
where $x=\cos\theta$ and the phases $\xi_\ell$ will be
discussed below, or as the sum
\begin{equation}
   M_{c_1} = \frac{\eta}{2p}\ \frac{1}{(1-x)^{1+i\eta}}\
                      e^{2i\Phi}\ F(k^2) \ .
   \label{eq:ff}
\end{equation}
The nuclear amplitude is given by
\begin{equation}
   M_N = \sum_{\ell=0}^\infty
                  (2\ell+1)\ e^{2i\sigma_{em,\ell}}\
                  \frac{e^{2i\delta_\ell}-1}{2ip}\, P_\ell(x) \ .
  \label{eq:nuc}
\end{equation}
Now there are choices that must be made.
Do we want to include in Eq.~(\ref{eq:ff}) a form factor $F(k^2)$,
or do we take $F \equiv 1$?
When we include a form factor, then we need to find a way to
calculate the corresponding Coulomb phases $\xi_\ell$.
Only in the case of point charges we know these Coulomb phases
analytically: $\xi_\ell=\sigma_\ell\equiv\Gamma(\ell+1+i\eta)$.
Because of the uncertainty of how to handle the long-range screening
of the Coulomb potential,
the overall phase $\Phi$ in Eq.~(\ref{eq:ff}) is really unknown.
There are two standard ways~\cite{Ta74,Se75,Ge76}
to treat this for point charges:
either we choose $\Phi=\sigma_0$, then $\xi_\ell=\sigma_\ell$,
or we choose $\Phi=0$, but then $\xi_\ell=\sigma_\ell-\sigma_0$.
In Nijmegen we opted for the latter choice. The $\sigma_{em,\ell}$
in Eq.~(\ref{eq:nuc}) is the sum of the choice for $\xi_\ell$ of
the standard Coulomb amplitude Eq.~(\ref{eq:coul}),
the two-photon phase~\cite{Au83} $\rho_\ell$, the
magnetic-moment phase~\cite{St90} $\phi_\ell$,
and the vacuum-polarization phase~\cite{Du57} $\tau_\ell$.

We hope that we made it clear that the phase of the Coulomb amplitude
is ``theorist dependent.'' The same holds then for the nuclear
amplitude. This means that therefore also the imaginary part
of the nuclear amplitude is theorist dependent.
This has repercussions for the optical theorem.
Consider the optical theorem in the case of purely elastic scattering.
When $f_\ell$ is the partial-wave nuclear scattering amplitude and
$\sigma^{(\ell)}_{elas}$ the hadronic cross section in this
partial wave, then
\begin{equation}
   \sigma^{(l)}_{elas} \sim | f_\ell |^2 =
      \sin^2 \delta_\ell \;\;\; {\rm and} \;\;\; {\rm Im}\:
      f_\ell = \sin\delta_\ell \sin(\delta_\ell+2\sigma_{em,\ell}) \ .
\end{equation}
Clearly, ${\rm Im}\,f_\ell \neq | f_\ell |^2$:
The optical theorem for these partial-wave amplitudes and
for the amplitude $M_N$ is obviously {\it not} valid.

Let us now return to the $pp$ forward dispersion relations.
One cannot write down a dispersion relation for the forward
scattering amplitude $M_N(x=1)$. First, one needs to correct
this amplitude for the electromagnetic interaction. This is
the same type of correction as for the $pp$ scattering length,
only it has to be done for {\it all} partial waves at {\it every}
energy.
These corrections are difficult to make and are often not very
acurate. Doing this with Gamow factors is certainly too inaccurate.
In the physical region below the pion-production threshold
the nuclear amplitude is constructed using a PWA.
In that PWA a choice was made for the $\pi N\!N$ coupling constant.
Above the pion-production threshold the treatment makes essential
use of the optical theorem.
But we did show that the optical theorem is not really valid for
$M_N$ nor for $M^{\,\prime}_N$.
In the physical region of the $\overline{N}\!N$ channel
one again needs the optical theorem.
The largest uncertainty comes from the treatment of the unphysical cut.
Many parameters need to be introduced in order that the fixed-$t$
dispersion relations can describe the experimental data with
sufficient accuracy.
This can only be done in a PWA of the $pp$ scattering data.
Such work could be performed, but it will require such an enormous
effort that it probably will not be done soon. Again such a
PWA requires as input a value for the $\pi N\!N$ coupling constant.
We therefore come to the conclusion that an accurate determination of
the $\pi^0 pp$ coupling constant in forward $pp$ dispersion relations
is not very realistic. It may not be impossible, but it
will require many man-years of hard and tedious work.
\vspace{\baselineskip}

\section{Conclusions}

We have seen that there is overwhelming evidence that the
charge-independent value for the $\pi N\!N$ coupling constant is
$f^2$ = 0.0750 with a total error of perhaps less than 9 in the last
digit. This error includes possible systematic errors.
We think that the present accuracy is such that the determination
of charge-independence breaking effects is soon within reach.

We would like to comment here on another, much-quoted result of the
Karlsruhe-Helsinki analysis~\cite{Ho75}: the value $\kappa_\rho=6.1$
of the vector-to-tensor ratio of the $\rho N\!N$ coupling constants.
Vector-meson dominance of the electromagnetic form factors gives
$\kappa_\rho=3.7$. The Nijmegen soft-core potential Nijm78~\cite{Na78}
produces $\kappa_\rho=4.2$; its update Nijm93~\cite{St94}
$\kappa_\rho=4.1$.
Because it is clear that the value of $\pi N\!N$ coupling constant
has changed a lot since 1975, one~\cite{Ma91} can
place also question marks by the more than 20 years old Karlsruhe
value for $\kappa_\rho$.
We may assume that the newer data will change that value.
\vspace{2\baselineskip}

\noindent {\Large\bf Acknowledgements}

\vspace{\baselineskip}

We would like to thank Tom Rijken for many discussions.
JJdeS would like to thank Greg Smith for the
invitation to the very pleasant MENU97 conference and for his support.
The support of Harold Fearing is also gratefully acknowledged.
MR would like to thank the Australian Research Council for their
financial support.
Part of this work was included in the research program of the
Stichting voor Fundamenteel Onderzoek der Materie (FOM) with
financial support from the Nederlandse Organisatie voor
Wetenschappelijk Onderzoek (NWO).

\pagebreak

\bibliographystyle{unsrt}

\end{document}